\documentclass[10pt,sigconf]{acmart}

\usepackage{amsmath, amssymb}
\usepackage{amsthm}
\usepackage{algorithm}
\usepackage{algorithmic}
\usepackage{color, array, colortbl}
\usepackage{graphicx}
\usepackage{multirow}

\usepackage{framed,color}
\definecolor{shadecolor}{rgb}{0.9,0.9,0.9}

\usepackage{wasysym}

\newcommand{\madalin}[1]{\todo[inline,author={Madalin}]{#1}}

\newcommand{\mathsym}[1]{{}}

\newcommand{\cancel}[1]{}

\newenvironment{mydesc}{\par}{\par}
\newcommand{\myitem}[1][*]{\par\textbf{#1}}

\begin{document}

\title[Security Landscape of Programmable Dataplanes]{Charting the Security Landscape of\\Programmable Dataplanes}

\author{Andrei-Alexandru Agape$^1$ \quad 
Madalin Claudiu Danceanu$^1$ \\
Ren{\'e} Rydhof Hansen$^1$ \quad 
Stefan Schmid$^2$\\
{\footnotesize $^1$ Aalborg University, Denmark}\quad 
{\footnotesize $^2$ University of Vienna, Austria}
}

\date{}

\settopmatter{printacmref=false, printccs=false, printfolios=true}

\renewcommand\footnotetextcopyrightpermission[1]{}
\setcopyright{none}

\settopmatter{printacmref=false, printccs=false, printfolios=true}

\acmDOI{}

\acmISBN{}
\acmPrice{}

\renewcommand{\shortauthors}{Agape et al.}

\sloppy

\begin{abstract}
Emerging programmable dataplanes will
revamp communication networks, allowing 
programmers to reconfigure and tailor switches
towards their need, in a protocol-independent manner.
While the community has articulated well
the benefits of such network architectures in terms of flexibility and 
performance, 
little is known today about the security implications.
We in this position paper argue that the programmable dataplanes
in general and P4 in particular introduce an uncharted security landscape.
In particular, we find that while some existing security studies 
on traditional OpenFlow-based networks still apply, P4 comes with
several specific components and aspects which change the attack surface
and introduce new challenges. We highlight several examples and
provide a first systematic security analysis.
\end{abstract}

\maketitle

\section{Introduction}

By outsourcing and consolidating the control over
network devices to a logically centralized controller
and by introducing open interfaces,
Software-Defined Networks (SDNs) in general and OpenFlow in particular
have enabled great flexibility
in how modern communication networks can be managed and operated.
However, while OpenFlow is a useful standard in that it allows to control 
switches from many different 
vendors in a unified manner, it is still ``fixed'' as it relies on the assumption that 
switches have a fixed well-known
behavior, as described in the data sheet of the switch ASIC;
moreover, the growing support for protocols combined with the
fact that OpenFlow only mandates the fields on the packets that one can match
upon, but not the actions to be performed after the match, makes it hardly scalable
and ambiguous~\cite{rep17}.

\begin{figure}[tp]
\centering
\includegraphics[width=.45\textwidth]{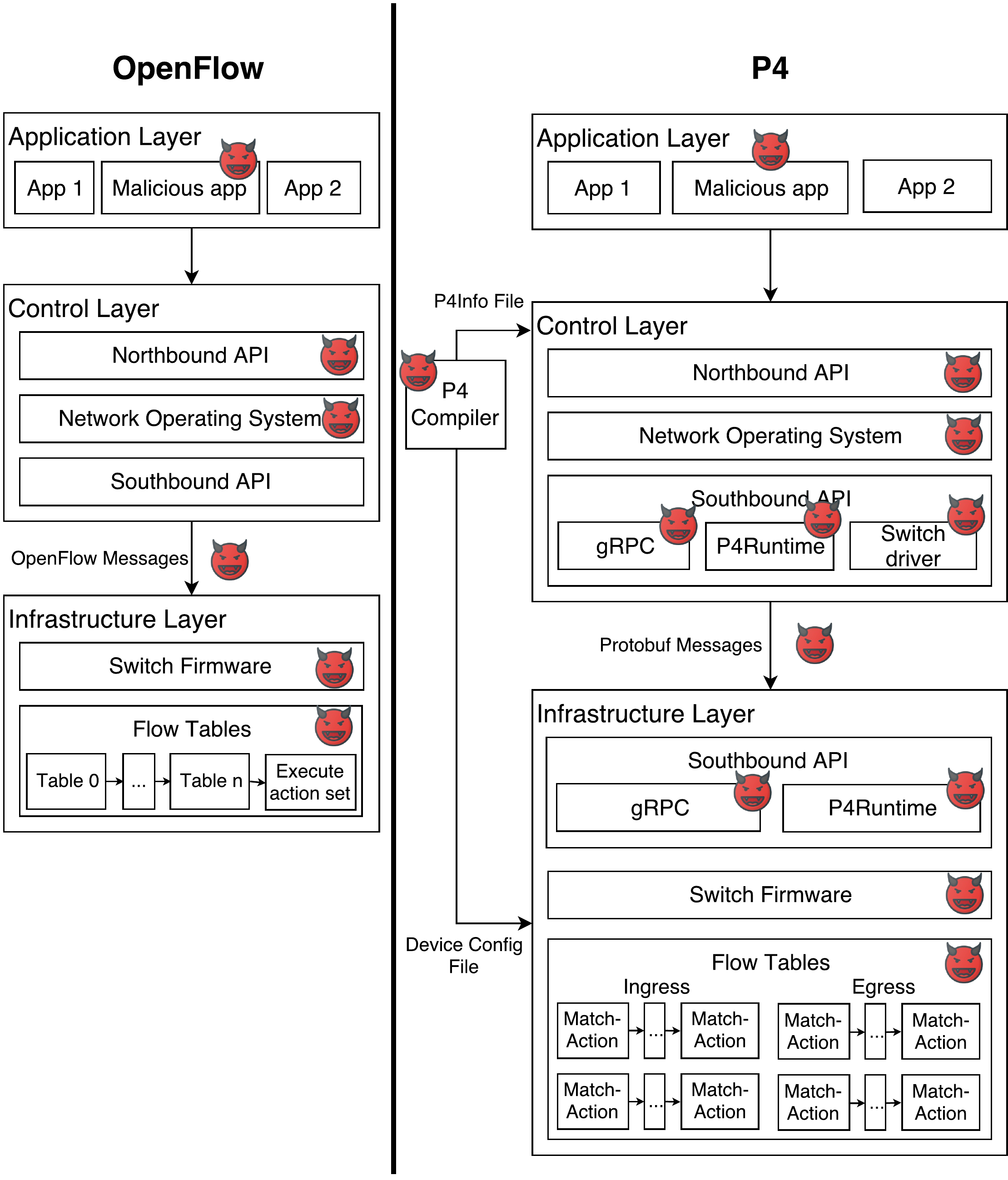}
\caption{OpenFlow vs P4 attack surface}\label{fig:of_vs_p4}
\end{figure}

\emph{Programmable dataplanes} and \emph{P4}~\cite{p4}
promise to fill this gap by offering an
open, flexible and silicon-independent API, 
reconfigurability (the way switches process packets 
can be changed at runtime),
protocol independence (switches are no longer tied to a 
specific network protocol), and target independence 
(packet processing functionality can be programmed 
independently of the specifics of the underlying hardware).
Besides a high-level programming language 
which can be compiled against many different types of execution machines
(called 
``P4 targets'', which have a P4 compiler back-end),
P4 offers a common API (called the \emph{P4 Runtime API}),
and allows to change and
immediately start using new forwarding tables, without
restarting the API or the control plane. 
The P4 language has no support 
for specific protocols, rather, the P4 programs are responsible for
specifying how a switch processes packets. 
%Once the programmer did that, 
These programs 
are then interpreted and processed by the compiled program
on the target device.

Our paper is motivated by the observation that programmable
dataplanes and P4 do not only
enable more flexible communication networks, interesting new use cases,
and an unprecedented performance, 
but also introduce a new attack surface and hence have 
implications on security.
Indeed, while the security of SDN architectures in general
and OpenFlow in particular have been explored in different
studies in the past, little is known about the security implications
of emerging P4 platforms.

\begin{figure}[tp]
\centering
\includegraphics[width=.45\textwidth]{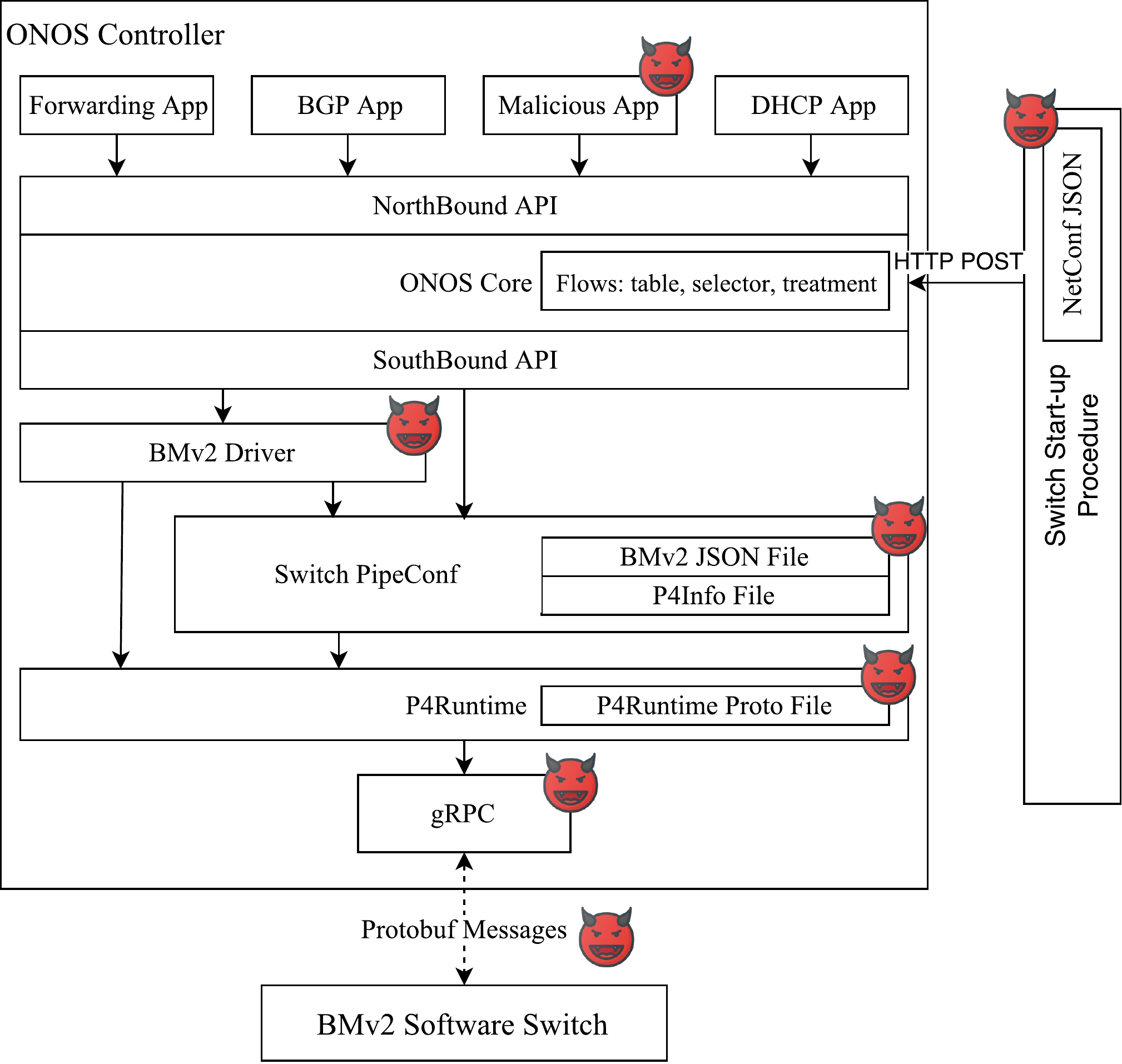}
\caption{Overview of P4 components, assets and attack points in ONOS Controller}\label{fig:overview-onos}
\end{figure}

\noindent \textbf{Our Contributions.}
This paper observes that programmable dataplanes and P4
change the security landscape, which so far is to
a large extent uncharted.  
Accordingly, we present 
a first systematic breakdown and approach
to study the attack surface and security implications
of emerging network
architectures supporting programmable 
packet forwarding.
Based on this breakdown, we characterize the 
possible attack surface of a P4-based SDN environment,
highlighting possible attacks and vulnerabilities related to 
the P4 language and
compiler, the controller (exemplified by ONOS), the 
P4 Runtime, as well as the switches (exemplified by the BMv2 switch).
Based on these insights, we discuss how specific
attacks and countermeasures can be implemented and report
on some experiments.
See Figure~\ref{fig:of_vs_p4} 
for an overview of the attack surface and comparison to traditional OpenFlow.

\section{Security Challenges}\label{sec:taxonomy}

\begin{figure}[tp]
\centering
\includegraphics[width=.45\textwidth]{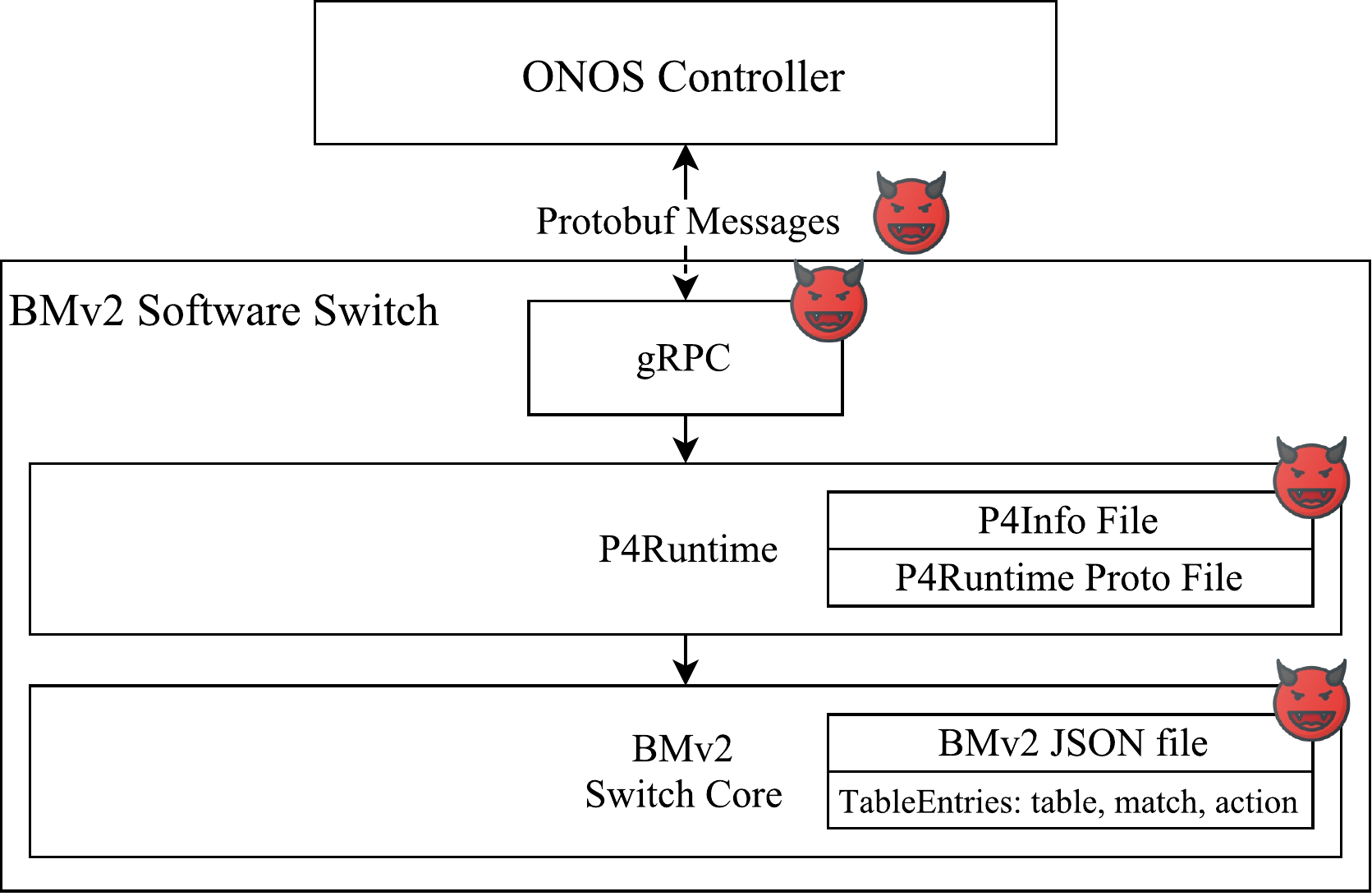}
\caption{Overview of P4 components, assets and attack points in BMv2 Switch}\label{fig:overview-bmv2}
\end{figure}

In this section we provide a brief overview of the main \emph{assets}
of a (typical) P4 SDN environment and a STRIDE analysis of the
\emph{attack surface} presented by such an environment.

\subsection{P4 Assets}

To perform our security and vulnerability analysis of P4, we
first have to identify and prioritise the potential targets, i.e., the
\emph{assets} of the P4 platform. An asset in this context is any
data, device, or other component that supports information related P4
activities. For convenience we have grouped the P4 assets of interest
into four general categories: control plane assets, channel plane
assets, dataplane assets, and the P4 compiler, see
Figure~\ref{fig:overview-onos} and Figure~\ref{fig:overview-bmv2} for an overview. In the following we briefly
describe the categories and their concomitant assets.

\noindent\textbf{Control Plane.}
The control plane, as a whole, is concerned with the routing process,
including ongoing management and setup of the process.

\begin{mydesc}
\myitem[Applications.] These are the primary assets in the control
  plane. An application is (potentially third-party) software designed
  to manage and perform specific actions within an SDN.
\myitem[P4Info.] This is the result of compiling the P4 program. This
  asset contains critical information such as tables, meters, counter,
  etc.\ as well as assigned IDs, enabling communication between
  controller and switch. This information is also used by both the P4
  controller to setup the forwarding configuration and the P4 runtime
  (denoted \emph{P4Runtime} in Figure~\ref{fig:overview-onos}) for
  translating IDs into objects.
\myitem[P4DeviceConfig.] The result of compiling the P4 control program
  to the target switch using the appropriate back-end compiler, e.g.,
  \emph{bmv2JSON} is the output of \emph{bmv2} back-end compiler. This
  asset is used by the controller, together with the \emph{P4Info}, to
  set up the forwarding plane configuration.
\myitem[SwitchPipeConf.] This is a controller application that defines
  the switch pipeline by using the \emph{P4Info} and
  \emph{P4DeviceConfig} assets to set up a mapping between P4 and
  platform specific objects.
\myitem[Switch driver.] The switch driver is a switch-specific
  application running on the controller and typically developed by the
  switch vendor. It provides an interface for adding and removing
  target specific table entries using the mapping set up by
  \emph{SwitchPipeConf}. As an example, in ONOS, it maps ONOS flows to
  P4 table entries.
\myitem[P4Runtime agent.] This is an application on the controller that
  serializes the P4 objects and the forwarding configuration to
  \emph{Protobuf} and calls the intended RPC methods.
\myitem[P4Runtime.proto.] The \emph{P4Runtime} protocol specification.
  It defines the RPC methods and messages that can be used between
  controller and switch. The protocol provides different RPC methods
  such as \emph{SetForwardingPipelingConfig} and \emph{StreamChanel}.
  On top of these, the protocol provides a multitude of message types.
  The \emph{P4Runtime.proto} is present in both controller and switch.
\myitem[gRPC.] The Remote Procedure Call (RPC) system developed at
  Google. It uses HTTP/2 for transport, \emph{Protocol Buffers} as the
  interface description language, and provides features such as
  authentication, bidirectional streaming and flow control,
  cancellation and timeouts etc.
\end{mydesc}
From the above, it should be clear that the control plane contains a
wide variety of assets, ranging from applications to simple files,
resulting in a wide attack surface.

\noindent\textbf{Channel Plane.}
The channel plane is concerned with inter-component communication
(through channels), mainly between controller and switch. For our use,
this plane comprises only a single asset:
\begin{mydesc}
\myitem[Protobuf messages.] These are the messages exchanged between the
  controller and switch, serialized using Protocol Buffers.
\end{mydesc}

\noindent\textbf{Dataplane.}
The dataplane is concerned with the actual forwarding of data
(packets) and shares (some) asset types with the control plane.
\begin{mydesc}
\myitem[gRPC.] This is similar to the \emph{gRPC} asset on the
  controller, with the difference that if the switch \emph{gRPC} is
  not available it is only the switch that cannot be controlled
  anymore, while if the controller gRPC is not available, the whole
  network becomes uncontrollable.
\myitem[P4Runtime.proto.] The P4Runtime protocol specification that
  defines the RPC methods and messages that can be used between
  controller and switch. Similar to the file present on the
  controller.
\myitem[Parser/de-parser.] These are defined in the P4 program as a
  deterministic finite automaton (DFA) using states and transitions.
\myitem[Flow tables.] These are the tables used to define exactly how
  the packets are forwarded and processed.
\end{mydesc}

\noindent\textbf{Compiler.}
The compiler is a key asset on the P4 platform: this is what enables
the rapid development of applications that can change major aspects of
a network. However, as with all programming, this also comes with many
potential risks requiring good (security aware) programming practices.
For our purposes, we consider different parts of the compiler as
separate assets.
\begin{mydesc}
\myitem[Front-end compiler.] This is a target-independent and standard
  part of the compiler that deals with the semantics checks, and that
  can be combined with a target-specific back-end to create a complete
  P4 compiler. We here take the front-end compiler to include various
  optimization passes, performed before the generated
  \emph{Intermediate Representation} (IR) is sent to the back-end
  compiler.
\myitem[Converter (P4-14 to P4-16).] This part of the compiler enables
  backward compatibility with the P4-14 version of the P4 language. It
  parses the P4-14 into version~1 of the Intermediate Language before
  it is converted to the \emph{Intermediate Representation} accepted
  by the back-end.
\myitem[Back-end compiler.] This is the main target-specific component
  of the compiler, usually developed by the vendor of the network
  components.
\end{mydesc}
With this we conclude the survery of the P4 assets we have identified.
It is possible to identify even more specific assets, but for an
initial mapping of major (potential) security vulnerabilities, we have
found that the above lists provide a good starting point.

\subsection{STRIDE Analysis and Attack Surface}

We now present a STRIDE analysis of the P4 platform,
based on the assets identified in the previous section.
STRIDE~\cite{stride} is a well-known model for categorising (potential) IT-security
threats and a useful tool for structuring threat-analysis of IT
systems. The name is a mnemonic derived from the threat categories
comprising the model: \emph{Spoofing}, \emph{Tampering},
\emph{Repudiation}, \emph{Information disclosure},
\emph{Denial-of-service}, and \emph{Evelation of privilege}. The
threat categories cover most, if not all, the ``classic''
threats/attacks that have been oberseved and reported in the
literature.

Since STRIDE analysis is fairly standard and well-known, we will not
discuss it in further detail here. We illustrate the general
methodology by briefly discussing an excerpt of the STRIDE
analysis for the \emph{P4Runtime} component (see
Table~\ref{fig:stride} for an overview).

\begin{table}[]
\centering
\caption{STRIDE analysis} \label{fig:stride}
{
\fontsize{6}{7}\selectfont
\bgroup
\def\arraystretch{1}
\begin{tabular}{|l|l|l|l|}
\hline
\textbf{Threat}                                                                     & \textbf{\begin{tabular}[c]{@{}l@{}}Property\\ violated\end{tabular}} & \textbf{Definition}                                                                                                   & \textbf{Example}                                                                                             \\ \hline
\multirow{3}{*}{Spoofing}                                                           & \multirow{3}{*}{Authentication}                                      & \multirow{3}{*}{\begin{tabular}[c]{@{}l@{}}Impersonating something\\ or someone else\end{tabular}}                    & \begin{tabular}[c]{@{}l@{}}Pretends to be another switch\\ in the network\end{tabular}                       \\ \cline{4-4} 
                                                                                    &                                                                      &                                                                                                                       & Pretends to be the controller                                                                                \\ \cline{4-4} 
                                                                                    &                                                                      &                                                                                                                       & \begin{tabular}[c]{@{}l@{}}Pretends to be the network\\ controller administrator\end{tabular}                \\ \hline
\multirow{2}{*}{Tampering}                                                          & \multirow{2}{*}{Integrity}                                           & \multirow{2}{*}{Modifying data or code}                                                                               & \begin{tabular}[c]{@{}l@{}}Intercept and modify\\ protobuf messages\end{tabular}                             \\ \cline{4-4} 
                                                                                    &                                                                      &                                                                                                                       & \begin{tabular}[c]{@{}l@{}}Take control of gRPC server\\ and modify the protobuf\\ messages\end{tabular}     \\ \hline
\multirow{2}{*}{Repudiation}                                                        & \multirow{2}{*}{Non-repudiation}                                     & \multirow{2}{*}{\begin{tabular}[c]{@{}l@{}}Claiming to have not\\ performed an action\end{tabular}}                   & \begin{tabular}[c]{@{}l@{}}A switch that does not follow\\ the controller instructions\end{tabular}          \\ \cline{4-4} 
                                                                                    &                                                                      &                                                                                                                       & \begin{tabular}[c]{@{}l@{}}A controller claiming that\\ a switch has not connected to it\end{tabular}        \\ \hline
\multirow{2}{*}{\begin{tabular}[c]{@{}l@{}}Information\\ Disclosure\end{tabular}}   & \multirow{2}{*}{Confidentiality}                                     & \multirow{2}{*}{\begin{tabular}[c]{@{}l@{}}Exposing information to\\ someone not authorized\\ to see it\end{tabular}} & \begin{tabular}[c]{@{}l@{}}Read device tables:\\ controller flows, \\switch tables\end{tabular}                \\ \cline{4-4} 
                                                                                    &                                                                      &                                                                                                                       & Read protobuf messages                                                                                       \\ \hline
\multirow{4}{*}{\begin{tabular}[c]{@{}l@{}}Denial of\\ Service\end{tabular}}        & \multirow{4}{*}{Availability}                                        & \multirow{4}{*}{\begin{tabular}[c]{@{}l@{}}Deny or degrade\\ service to users\end{tabular}}                           & \begin{tabular}[c]{@{}l@{}}Crashing the P4Runtime\\ gRPC service\end{tabular}                                \\ \cline{4-4} 
                                                                                    &                                                                      &                                                                                                                       & \begin{tabular}[c]{@{}l@{}}Flooding the switch-\\ controller channel\end{tabular}                            \\ \cline{4-4} 
                                                                                    &                                                                      &                                                                                                                       & \begin{tabular}[c]{@{}l@{}}Modify and invalidate\\ protobuf messages\end{tabular}                            \\ \cline{4-4} 
                                                                                    &                                                                      &                                                                                                                       & \begin{tabular}[c]{@{}l@{}}Intercept and deny arrival\\ of the packets to the intended\\ device\end{tabular} \\ \hline
\multirow{2}{*}{\begin{tabular}[c]{@{}l@{}}Elevation\\ of\\ privilege\end{tabular}} & \multirow{2}{*}{Authorization}                                       & \multirow{2}{*}{\begin{tabular}[c]{@{}l@{}}Gain capabilities without\\ proper authorization\end{tabular}}             & \begin{tabular}[c]{@{}l@{}}A switch changing\\ information in the controller\end{tabular}                    \\ \cline{4-4} 
                                                                                    &                                                                      &                                                                                                                       & \begin{tabular}[c]{@{}l@{}}Configure a switch and\\ decide how the traffic is handled\end{tabular}           \\ \hline
\end{tabular}
\egroup
}
\end{table}

\textbf{Spoofing.} A potential spoofing attack would be an attacker
  (successfully) masquerading as a (different) switch in the network.
  This would allow the attacker to elicit information from the
  controller, such as de-/parser configuration, pipeline
  configuration, forwarding table entries, and how table-miss flows
  are handled.

\textbf{Tampering.} If an attacker can modify, i.e., tamper with,
  protobuf messages can violate both confidentialiy, integrity, and
  availability properties of the network. An attacker that can take
  control of the switch \emph{gRPC} or the controller \emph{gRPC}, can
  modify the sent and received protobuf messages, thus controlling the
  switch or the entire network.
  
\textbf{Repudiation.} In a \emph{repudiation} attack, an attacker can
  make the switch refuse configurations from controller and claim they
  were not received, thus making the switch uncontrollable. An
  attacker can make the controller refuse connections from switches
  that try to connect to it and claim that no connection were
  instantiated, rendering the switch unable to handle traffic.
  
\textbf{Information disclosure.} An attacker with a presence on the
  network may be able to pick up information that is either sent in
  the clear, such as unencrypted protobuf messages, messages picked up
  directly from the control plane, or even exploiting specific timing
  properties for an advanced timing-attack on the controller or the
  switches.
  
\textbf{Denial-of-service.} A \emph{denial-of-service attack} may crash
  the \emph{gRPC} service, making the communication between the switch
  and the controller unavailable, potentially wrecking havoc in the
  network.
  
\textbf{Elevation of privilege.} As part of an \emph{elevation of
    privilege}, an attacker can write malicious applications, and
  based on the controller configuration, allow it to read, modify or
  deny data and services. It may also modify the forwarding tables.

To obtain a high-level overview of a system's security
stance, it is often useful to consider the system's \emph{attack
  surface}. In general, the attack surface of a system is the sum of
the different points (``attack vectors'') where an unauthorized user
(``attacker'') can try to enter data to or extract data from an
environment~\cite{attack-surface-metric}. 
Here, the attack surface is composed of: (1)~the data in
the system and messages exchanged, (2)~the methods for processing
applications, e.g., request/response methods, (3)~the communication
channels, e.g., HTTP, TCP.

\section{P4 Language and Compiler}

The main components of the P4 language compiler are the parser (either
P4-14 or P4-16), an IR converter that enables backwards compatibility with the
P4-14 language, a fixed front-end component, customizable mid-ends, and back-ends
provided by the vendor for specific targets.
Since the front-end and mid-end are standard and mostly fixed, bugs and/or vulnerabilities
at this level can generate incorrect IR for all back-ends that are using them.
Even though an attacker may not be able to directly modify the P4 programs, the
source code of the front-end compiler and the language specifications can be found
on the official websites and repositories.

\textbf{P4 compiler fuzzer.} One option the attacker has is to write P4 programs that can either crash the
compiler, generate invalid target code or target code that is not following
the original intentions. In order to automatically generate random test-cases, a
compiler fuzzer~\cite{compfuzz} can be implemented.

\textbf{Race conditions.} Another interesting and novel attack may
involve the use of \emph{concurrency}, i.e.,
an attacker may try to find out whether the race
conditions are handled correctly.
In fact, \emph{the extern blocks instantiated by a P4 program
are global, and shared across all threads; 
if extern blocks mediate access to state (e.g.,
counters, registers) [$\ldots$] these
stateful operations are subject to data races.}~\cite{p4lang-spec}.
There are tools for the automatic detection of race conditions,
such as \emph{RaceMob},
\emph{RaceFuzzer} and \emph{RaceChecker}, 
however, currently none exist for P4.

\textbf{P4-14 to P4-16 converter.} Because of its backward compatibility with P4-14, 
the P4C introduces another point of
attack in the sense of the P4-14 to P4-16 converter. 
Bugs or vulnerabilities in the
converter can introduce another class of problems
 related to the P4-14 version of P4
programs.
Using a compiler fuzzer 
to generate P4-14 programs can be
one way to find bugs in the converter. Other vulnerabilities can be related to the
converter parallel actions. An example of such a bug has been reported to the official
repository (Issue 246) and it is still not
clear if it was solved due to the P4 1.1 language specification being ambiguous.

\textbf{P4 language benchmark.} As presented earlier the back-end compiler is target specific, 
therefore multiple compilers
are being developed without a clearly defined standard. 
While the benchmark proposition could help creating
a standard for the P4 compilers, it should be built and adopted by the whole community, which e.g., is not the case of Whippersnapper~\cite{whippersnapper}, 
or at least not yet. Currently
four sample backends are available on the P4Lang repository: p4c-bm2-ss, p4c-ebpf,
p4test and p4c-graphs; the latter two are used for debugging and generating graphs
of top-level control flows.

\textbf{Undefined behaviour.} One possible attacking point is represented by the P4 language specifications,
more precisely the ``Undefined behavior'' chapter. 
As stated in this chapter ``there
are a few places where evaluating a P4 program can result in undefined behaviors:
out parameters, uninitialized variables, accessing header fields of invalid headers, and
accessing header stacks with an out of bounds index''.
An attacker can find code patterns that have a vulnerability potential by making
use of the undefined behavior. In this sense, the new programmable switches have
exploitability potential if the undefined behaviour is handled differently by some
compilers compared to others.

\textbf{ASSERT-P4.} Assertion-based verification~\cite{assertp4} can be used 
to check general security and  correctness
properties of P4 programs. 
From an attacker point of view, the tool can also 
be used to find vulnerabilities in open-source
P4 applications.
ASSERT-P4 is a tool designed in this sense that can be used to annotate the P4 programs with assertions, translate it to a C-based model and verify it using a symbolic execution engine. The engine tests all possible paths and reports any assertion failure.

Table~\ref{fig:attacks} and Table~\ref{fig:countermeasures} summarizes the attacks 
and countermeasures regarding the P4 language and compiler.

\section{Case Study: ONOS and BMV2}

To illustrate our general security analysis 
 and make it more concrete, we here consider three specific
examples from the ONOS project, namely the ONOS controller, the ONOS
P4 Runtime, and the BMv2 Switch.

\subsection{The ONOS Controller}

Analysing the ONOS controller is particularly interesting, as it is
currently also modified to support P4. The ONOS controller runs on top
of a Java Virtual Machine and is mainly used to alter the tables
entries on the controlled devices in the network in order to properly
manage the network during its operation.

\textbf{Malicious applications.} Malicious applications can disclose confidential information, slow
down the service or even crash the controller. ONOS runs as a single
process including all its internal components and also the installed
and active applications. Consequently, an attacker can crash the
controller by simply closing that process, which would not only close
the application but the entire ONOS controller.

\textbf{Controller configuration.} Another type of attack could exploit the fact that networks 
are changing continuously, with devices being removed,
added, or crash. 
The controller has to
be configured properly and have the right applications 
running in order to support such churn.
The complexity of the required configurations
(related to encryption, usage of strong credentials, 
only activating the
features that are required for the specific network, and so on)
may introduce errors and increase the attack surface. 
One example of such a configuration is that ONOS Command Line Interface (CLI) has a default insecure client (Apache Karaf) since it relies on a well-known private key. Installing the security countermeasure (onos-secure-ssh) as described in the documentation is not working due to it being outdated. The fact that some versions of ONOS have the Security-mode disabled by default, and enabling it can be a tedious process, contributes also to the overall security.

\textbf{Authentication.} Access to the CLI, 
Web GUI, or the REST API is done through authentication. 
While the ONOS CLI uses
public/private key authentication, the GUI and the REST API 
require username/password credentials.
The brute-force type of attacks exploit the fact that users
often use simple enough combinations of username and
password.

Table~\ref{fig:attacks} and Table~\ref{fig:countermeasures} summarizes the 
attacks and countermeasures regarding the controller.

\subsection{The P4 Runtime}

We identify two main vulnerabilities
which regard the P4 Runtime (both
the Java ONOS controller and the C++ BMv2 switch): a man-in-the-middle
attack and channel flooding.

\textbf{Man-in-the-middle.} A man-in-the-middle attack requires that the 
adversary has access to the channel
through which important packets travel. 
This type of attack is particularly relevant in the context of 
P4 programmable switches with
P4Runtime support because the gRPC messages communicated
on the channel between
the switch and the controller, containing much sensitive information:
e.g., switch configuration files, the tables
available, and other control messages altering the table entries.
The information captured by the adversary can be used for other types
of attacks such as spoofing. Also worth mentioning is the fact that 
if the switch sends the packets that do not match any tables to the
controller, the adversary 
is also able to capture
these messages which could contain sensitive information such as credentials or other
personal information.
Even though the messages are serialized into binaries 
using protobuf, they are not encrypted and can be deserialized 
by using the protobuf compiler. The deserialization
process requires the P4Runtime protocol specification for 
protobuf and the
P4 program information file containing the \textit{ID}s for 
tables and other P4 objects. 
The P4Runtime protocol specification can be easily 
obtained since it is publicly available while the P4Info file 
can be obtained by listening for the initial start-up process
of the switch when this file is transmitted from the controller to the switch.

\textbf{Channel flooding.} In the context
of P4 programmable switches controlled using the P4Runtime protocol 
there is a
single P4Runtime agent in the controller while 
each switch has its own P4Runtime
agent. Thus flooding the channel with packets from 
one or multiple switches in the
network can lead to slower response time or
 even denial of service of the P4Runtime
agent in the controller.
The attack can be conducted in both directions: e.g.,
the controller, probably
through a rogue application, may flood the switches with 
many control messages in order
to affect the behavior and response time of the network.

Tables~\ref{fig:attacks} and~\ref{fig:countermeasures} summarize the 
attacks and countermeasures regarding the 
P4 runtime.

\subsection{The BMv2 Switch}

There are also potential vulnerabilities on the
switch side. 
For example, the BMv2 Switch with P4Runtime support has 
its tables populated and altered
by the controller through remote procedure calls 
with messages serialized using the
protobuf protocol. This fact can be exploited by an adversary 
by sending control messages to the switch as if these messages 
originate from the controller.
In order for the adversary to be able to perform such 
an attack, it needs to be able to serialize messages 
using the protobuf protocol for sending the
information wanted to the switch. For serializing the data 
the attacker needs to
posses some knowledge regarding the switch such as the description 
of the tables
together with their assigned ids which are contained in the P4Info file as well as the
switch behavior description which is located in the compiled 
BMv2 JSON config
file. 
The attacker can obtain this information by 
using other attacks such as the
man-in-the-middle attack.
By spoofing the controller, an attacker can alter 
the table entries in the switches
completely changing the behavior of the devices. 
This can be used to bypass firewalls
or change the configuration of the devices in a such a way that would somehow
benefit the adversary.
Table~\ref{fig:attacks} and Table~\ref{fig:countermeasures} summarizes the 
attacks and countermeasures regarding the switch.

\begin{table}[]
\centering
\caption{Attacks}
\caption*{\tiny Legend: \Circle = Low, \LEFTcircle = Medium, \CIRCLE = High}
\label{fig:attacks}
{
\fontsize{7}{9}\selectfont
\begin{tabular}{|l|l|c|c|}
\hline
Name                                                                                                 & Component  & Impact & Difficulty \\ \hline
\begin{tabular}[c]{@{}l@{}}Exploiting Undefined \\ Behaviour\end{tabular} & Compiler   & \LEFTcircle  & \LEFTcircle     \\ \hline
Exploiting Concurrency                                                                               & Compiler   & \LEFTcircle & \LEFTcircle     \\ \hline
\begin{tabular}[c]{@{}l@{}}Assertion based\\ verification\end{tabular}    & Compiler   & \LEFTcircle & \CIRCLE       \\ \hline
Malicious application                                                                                & Controller & \CIRCLE   & \LEFTcircle     \\ \hline
\begin{tabular}[c]{@{}l@{}}Exploiting \\ misconfiguration\end{tabular}    & Controller & \CIRCLE   & \LEFTcircle     \\ \hline
Login brute-force                                                                                    & Controller & \CIRCLE   & \LEFTcircle     \\ \hline
Man-in-the-Middle        & P4 Runtime & \CIRCLE   & \LEFTcircle     \\ \hline
Channel flooding                                                                                     & P4 Runtime & \CIRCLE   & \LEFTcircle     \\ \hline
Spoofing the controller                                                                              & Switch     & \CIRCLE   & \CIRCLE       \\ \hline
\end{tabular}
}
\end{table}

\begin{table}[]
\centering
\caption{Countermeasures}
\caption*{\tiny Legend: \Circle = Low, \LEFTcircle = Medium, \CIRCLE = High}
\label{fig:countermeasures}
{
\fontsize{7}{9}\selectfont
\begin{tabular}{|l|l|c|c|}
\hline
Name                                                                                               & Component      & Impact & Difficulty \\ \hline
\begin{tabular}[c]{@{}l@{}}Assertion based\\ verification\end{tabular} & Compiler       & \LEFTcircle & \CIRCLE       \\ \hline
Compiler fuzzer                                                                                    & Compiler       & \CIRCLE   & \CIRCLE       \\ \hline
\begin{tabular}[c]{@{}l@{}}Whippersnapper \\ benchmark\end{tabular}     & Compiler       & \Circle    & \LEFTcircle     \\ \hline
ONOS secure mode                                                                                   & Controller     & \CIRCLE   & \Circle        \\ \hline
\begin{tabular}[c]{@{}l@{}}Symbolic Execution \\ Analysis\end{tabular}  & Controller     & \LEFTcircle & \CIRCLE       \\ \hline
Brute-foce protection                                                                              & Controller     & \LEFTcircle & \Circle        \\ \hline
Securing the channel                                                                               & P4 Runtime     & \CIRCLE   & \Circle        \\ \hline
Reactive firewall application                                                                      & P4 Runtime     & \CIRCLE   & \LEFTcircle     \\ \hline
\end{tabular}
}
\end{table}

\section{Conclusion}\label{sec:conclusion}

One may argue that at least the security of SDNs and OpenFlow
is fairly well-understood today, and indeed, many existing known weaknesses and
vulnerabilities, as well as countermeasures known from SDN architectures
in general also directly apply to P4.
Yet, in this paper we have shown that P4 architectures in general and
programmable dataplanes in particular come with many specific
properties that have the potential to change the security
landscape and, as we argue, require special attention. 

{
\bibliographystyle{ACM-Reference-Format}
\bibliography{p4sec}
}

\end{document}